\documentclass[12pt]{article}

\begin{document}
\begin{titlepage}
\vbox{}
\vspace{1.5cm}
\begin{center}{\Large \bf COMPLETE POSITIVITY\\
\vspace{.2cm}
AND CORRELATED NEUTRAL KAONS}
\end{center}
\vspace{1.0cm}
\vspace{2cm}
\begin{center}
{\bf F. Benatti}\\
Dipartimento di Fisica Teorica, Universit\`a di Trieste\\
Strada Costiera 11, 34014 Trieste, Italy\\
and\\
Istituto Nazionale di Fisica Nucleare, Sezione di Trieste\\
\vspace{1cm}
{\bf R. Floreanini}\\
Istituto Nazionale di Fisica Nucleare, Sezione di Trieste\\
Dipartimento di Fisica Teorica, Universit\`a di Trieste\\
Strada Costiera 11, 34014 Trieste, Italy
\end{center}

\vspace{3cm}

\centerline{\bf Abstract}
\medskip
\narrower\narrower
\noindent
In relation with experiments on correlated kaons at
$\phi$-factories, it is shown that the request of complete positivity is
necessary in any physically consistent description of neutral kaons as
open quantum systems.

\end{titlepage}

\vfill\eject

Non-standard phenomenological models incorporating loss of 
quantum coherence and entropy increase can be studied 
in the broad framework of open quantum systems.\cite{1}-\cite{3}
These systems can be modelled
as being small subsystems in ``weak'' interaction with large environments.
The reduced dynamics for the subsystem is obtained 
by eliminating, {\it i.e.} tracing over, the environment degrees of freedom;
in the markovian approximation, justified in most physical contexts,
it consists of a set of completely positive linear transformations,
with forward in time composition law (semigroup property) and entropy
increase. These maps form a so-called quantum dynamical semigroup.

This description is rather general and has been used to model
a large variety of physical situations, ranging from the study
of quantum statistical systems,\cite{1}--\cite{3} 
the analysis of dissipative effects
in quantum optics,\cite{4}--\cite{6} the treatment of 
the interaction of a microsystem
with a macrosystem.\cite{7}--\cite{9} 
It has also been recently applied to the
analysis of the evolution and decay of the system of neutral
kaons.\cite{10}--\cite{14} 
(For motivations and earlier results, see \cite{15}--\cite{22}).

According to standard phenomenology,\cite{23} kaon states can be effectively
described by $2\times 2$ density matrices $\rho$ acting on a two dimensional 
Hilbert space.
The time evolution $\omega_t$ will transform the initial state $\rho$ into the
state $\rho(t)\equiv\omega_t[\rho]$, at time $t$.
A consistent statistical description of the initial density
matrix $\rho$ as a state is assured by
the positivity of its eigenvalues that are interpreted as probabilities.
Clearly, for this description to hold for all times,
the evolution map $\omega_t$ must be positive, {\it i.e.} it
must preserve the positivity of the eigenvalues of
$\rho(t)$, for any $t$.

Complete positivity\cite{1}--\cite{3} is a more stringent condition;
it guarantees the positivity of the eigenvalues of density matrices
describing states of correlated kaons, {\it e.g.} those produced in 
$\phi$-meson decays.
States of entangled, but not dynamically interacting kaons, 
evolve according to the factorized product $\omega_t\otimes\omega_t$
of the single-kaon dynamical maps.\cite{12}
If $\omega_t$ is not completely positive, there are instances
of correlated states that develop negative eigenvalues; in such
cases, their statistical and physical interpretation is lost.\cite{13}
The aim of this note is to show that the issue of
complete positivity is not only theoretical, but can be given
experimental relevance. Indeed, we shall explicitly find that
some experimentally accessible kaon observables, defined to be positive,
would return, in absence of complete positivity, negative mean
values.

The evolution in time of any kaon state $\rho$ can be described in general
by an equation of the following form:
\begin{equation}
\label{1}
{\partial\rho(t)\over\partial t}=-iH_{\rm eff}\, \rho(t)+i\rho(t)\, 
H_{\rm eff}^\dagger +L_D[\rho(t)]\ .
\end{equation}
The first two pieces in the r.h.s. give the standard hamiltonian
contribution, while $L_D$ is a linear map that encodes possible 
dissipative, non-standard effects.

The effective hamiltonian,
\begin{equation}
\label{2}
H_{\rm eff}=M-i{\mit{\Gamma}}/2\ ,
\end{equation}
includes a non-hermitian part
describing the kaon decay; its generic form further encodes possible
indirect $CP$ and $CPT$ violations.
It is convenient to introduce the even and odd 
$CP$ eigenstates $|K_1\rangle$,
$|K_2\rangle$ and the eigenstates of $H_{\rm eff}$
\begin{equation}
\label{3}
|K_S\rangle=\frac{|K_1\rangle+\epsilon_S|K_2\rangle}
{\sqrt{1+|\epsilon_L|^2}}\ ,\
|K_L\rangle=\frac{\epsilon_L|K_1\rangle+|K_2\rangle}
{\sqrt{1+|\epsilon_S|^2}}\ ,
\end{equation}
such that $H_{\rm eff}|K_S\rangle=\lambda_S|K_S\rangle$ and 
$H_{\rm eff}|K_L\rangle=\lambda_L|K_L\rangle$.
The entries of the matrices $M$ and 
$\mit\Gamma$ can be expressed in terms of the two complex parameters 
$\epsilon_S$, $\epsilon_L$ and the $K_S$, $K_L$ masses,
$m_S$, $m_L$, and decay widths $\gamma_S$, $\gamma_L$, so that 
$\lambda_S=m_S-i\gamma_S/2$, $\lambda_L=m_L-i\gamma_L/2$.
For later use, we introduce the following positive combinations:
$\Delta\Gamma=\gamma_S-\gamma_L$, $\Delta m=m_L-m_S$,
as well as the complex quantities 
$\Gamma_\pm=\Gamma\pm i \Delta m$ and 
$\Delta\Gamma_\pm=\Delta\Gamma\pm 2i\Delta m$, with 
$\Gamma=(\gamma_S+\gamma_L)/2$.

The form of the additional piece $L_D[\rho]$ is uniquely fixed by the 
physical requirements that the complete time-evolution
$\omega_t$ needs to satisfy;
as already mentioned in the introductory remarks, these are
semigroup property, entropy increase and complete positivity.
Expressing the kaon state $\rho$ as the $4$-vector 
$(\rho_0,\rho_1,\rho_2,\rho_3)$ along the basis consisting of the identity 
matrix $\sigma_0$ and the Pauli matrices $\sigma_i$, $i=1,2,3$, 
the additional linear piece $L_D$ acts as the $4\times 4$ matrix:\cite{10}
\begin{equation}
\label{4}
[L_D]=-2\pmatrix{
0&0&0&0\cr
0&a&b&c\cr
0&b&\alpha&\beta\cr
0&c&\beta&\gamma},
\end{equation}
where $a$, $b$, $c$, $\alpha$, $\beta$ and
$\gamma$ are six real parameters, with $a$, $\alpha$ and $\gamma$ non
negative. Further, these parameters must satisfy the following
inequalities
\begin{eqnarray}
\label{5}
&&2R\equiv\alpha+\gamma-a\geq0\ ,\qquad RS\geq b^2\ ,\\
\label{6}
&&2S\equiv a+\gamma-\alpha\geq0\ ,\qquad RT\geq c^2\ ,\\
\label{7}
&&2T\equiv a+\alpha-\gamma\geq0\ , \qquad ST\geq\beta^2\ ,\\
\label{8}
&&RST\geq 2\, bc\beta+R\beta^2+S c^2+T b^2\ ,
\end{eqnarray}
which are necessary and sufficient conditions for the complete positivity of
the time-evolution map $\omega_t$ generated by~(\ref{1}).

It should be stressed that in absence of the piece $L_D[\rho]$,
pure states ({\it i.e.} states of the form $|\psi\rangle\langle\psi|$)
are transformed by the evolution equation (\ref{1}) back into pure states,
even though probability is not conserved, a direct consequence
of the presence of a non-hermitian part in the effective hamiltonian
$H_{\rm eff}$. Only when the extra piece $L_D[\rho]$ is also present, 
$\rho(t)$ becomes less ordered in time due to a mixing-enhancing mechanism:
it produces dissipation and possible loss of quantum coherence.%
\footnote{Although in principle 
these phenomena could also contribute to the decay 
process of the $K$-mesons, their effects can be estimated
to be negligible for any practical considerations;\cite{9} the description
of the $K$-meson ``self-dynamics'' by means of the effective
hamiltonian (\ref{2}) is therefore appropriate.}

A rough dimensional estimate gives the magnitude 
of the non-standard parameters in (\ref{4}); 
they are at most 
of the order $m_K^2/M_P\simeq 10^{-19}$ GeV, with $m_K$ the kaon mass
and $M_P$ the Planck mass;\cite{16}, \cite{21}, \cite{10}, \cite{14} 
the magnitude of this estimate is not 
far from the value 
of $\Delta\Gamma\,\epsilon_S$ and $\Delta\Gamma\,\epsilon_L$.
This allows a perturbative treatment 
of the evolution equation in (\ref{1}).
It is thus possible to compare the predictions deduced from the
solutions $\rho(t)$ of (\ref{1}) 
with the available data of various observables accessible in kaon
experiments.\cite{11}
Unfortunately, the accuracy of the present data, 
essentially from fixed target experiments, are not sufficient to draw a 
definite conclusion whether the constraints (\ref{5})--(\ref{8}) 
are obeyed or not.
More precise results should come from the 
experiments on correlated kaons in
the so-called $\phi$-factories.\cite{24}

Since the $\phi$-meson has spin 1, the two neutral spinless kaons produced
in a $\phi$-decay fly apart
with opposite momenta in the meson $\phi$ rest-frame; their
corresponding density matrix $\rho_A$
is antisymmetric in the spatial labels.
By means of the projectors onto the $CP$ eigenstates,
\begin{equation}
\label{9}
P_1=|K_1\rangle\langle K_1|\ ,\qquad P_2=|K_2\rangle\langle K_2|\ ,
\end{equation}
and of the off-diagonal 
operators, 
\begin{equation}
\label{10}
P_3=|K_1\rangle\langle K_2|\ ,\qquad 
P_4=|K_2\rangle\langle K_3|\ , 
\end{equation}
we can write
\begin{equation}
\label{11}
\rho_A={1\over 2}(P_1\otimes P_2 + P_2\otimes P_1
- P_3\otimes P_4 - P_4\otimes P_3)\ .
\end{equation}

The time evolution of a system of two correlated neutral $K$-mesons,
initially described by $\rho_A$, can be analyzed using the sigle
$K$-meson dynamics so-far discussed. Indeed, as in ordinary quantum
mechanics, it is natural to assume that, once
produced in a $\phi$ decay, the kaons evolve in time each according to the
completely positive map $\omega_t$ generated by (\ref{1}).
As already remarked, this assures that the resulting total evolution map 
is still positive for any time and of semigroup type.
We stress that this choice is the only natural possibility if one
requires that after tracing over the degrees of freedom of one particle, 
the resulting dynamics for the remaining one be completely positive, 
of semigroup type and independent from 
the initial state of the other particle.

Within this framework, the density matrix that describes a situation in which
the first $K$-meson has evolved up to proper time $t_1$ 
and the second up to proper time $t_2$ is given by:
\begin{eqnarray}
\nonumber
&&\rho_A(t_1,t_2)\equiv
\big(\omega_{t_1}\otimes\omega_{t_2}\big)\big[\rho_A\big]\\
\label{12}
&&={1\over 2}\Big[P_1(t_1)\otimes P_2(t_2)\ 
+\ P_2(t_1)\otimes P_1(t_2)\ 
- P_3(t_1)\otimes P_4(t_2)-P_4(t_1)\otimes P_3(t_2)\Big]
\end{eqnarray}
where $P_i(t_1)$ and $P_i(t_2)$, $i=1,2,3,4$, represent the evolution
according to (\ref{1}) of the initial operators (\ref{9}), (\ref{10}), 
up to the time $t_1$ and $t_2$, respectively.
In the following, we shall set $t_1=t_2=t$, and simply call
$\rho_A(t)\equiv\rho_A(t,t)$ the evolution of (\ref{11}) up to
proper time $t$.

The statistical description of $\rho_A(t)$ allows us to give a meaningful
interpretation of the quantities
\begin{equation}
\label{13}
{\cal P}_{ij}(t)={\rm Tr}[\rho_A(t)P_i\otimes P_j]\ ,\quad i,j=1,2\ ,
\end{equation}
as the probabilities to have one kaon in the
state $|K_i\rangle$ at proper time $t$, while the other
is in the state $|K_j\rangle$ at the same proper time.
When $i,j=3,4$, the quantities ${\cal P}_{ij}(t)$ are complex and 
do not represent directly joint probabilities.
However, as we shall see, they can still be obtained from data
accessible to experiments.

Following the discussion in \cite{12}, up to first order 
in all small parameters, one finds:
\begin{eqnarray}
\label{14}
{\cal P}_{11}(t)&=&{\gamma\over\Delta\Gamma}\, e^{-2\Gamma t}
\Bigl(1-e^{-\Delta\Gamma t}\Bigr)\ ,\\
\label{15}
{\cal P}_{12}(t)&=&{e^{-2\Gamma t}\over 2}\ ,\\ 
\label{16}
{\cal P}_{13}(t)&=&2\,e^{-2\Gamma t}\,{c+i\beta\over\Delta\Gamma_+}\,
\Bigl(1-e^{-t\Delta\Gamma_+/2}\Bigr)\ ,\\
\label{17}
{\cal P}_{22}(t)&=&{\gamma\over\Delta\Gamma}\,e^{-2\Gamma t}
\Bigl(e^{\Delta\Gamma t}-1\Bigr)\ ,\\
\label{18}
{\cal P}_{23}(t)&=&2\,e^{-2\Gamma t}\,{c+i\beta\over\Delta\Gamma_-}\,
\Bigl(1-e^{t\Delta\Gamma_-/2}\Bigr)\ ,\\
\label{19}
{\cal P}_{33}(t)&=&e^{-2\Gamma t}\,{2b+i(\alpha-a)\over2\Delta m}\,
\Bigl(1-e^{-2i t\Delta m}\Bigr)\ ,\\
\label{20}
{\cal P}_{34}(t)&=&-{e^{-2\Gamma t}\over 2}
\Bigl(1-2(\alpha+a-\gamma)t\Bigl)\ .
\end{eqnarray}
The remaining quantities ${\cal P}_{ij}(t)$ can be derived from the
previous expressions by using the following properties:
\begin{eqnarray}
\label{21}
&&{\cal P}_{ij}(t)={\cal P}_{ji}(t)\ , \qquad i,j=1,2,3,4\ ,\\
\label{22}
&&{\cal P}_{i3}(t)={\cal P}_{i4}^*(t)\ , \qquad i=1,2\ ,\\
\label{23}
&&{\cal P}_{44}(t)={\cal P}_{33}^*(t)\ .
\end{eqnarray}
Putting $a=b=c=\alpha=\beta=\gamma=0$, one obtains the standard  
quantum mechanical effective description that
evidentiates the singlet-like anti-correlation in $\rho_A(t)$:
${\cal P}_{ii}(t)\equiv 0$.

We emphasize that none of the above expressions contain the standard $CP$, 
$CPT$-violating parameters $\epsilon_S$, $\epsilon_L$. This fact 
makes possible, at least in principle, 
a direct determination of the non-standard parameters
irrespectively of the values of $\epsilon_S$, $\epsilon_L$;
one needs to fit the previous expressions of 
the quantities ${\cal P}_{ij}(t)$ with actual data from
experiments at $\phi$-factories. 

To be more specific, we shall now explicitly show how the 
quantities ${\cal P}_{ij}$ can be directly related
to frequency countings of decay events.
First, notice that, given any single-kaon time-evolution
$\rho\mapsto\rho(t)$, the matrix elements of the state
$\rho(t)$ at time $t$ can be measured by identifying appropriate
orthogonal bases in the two-dimensional single kaon Hilbert space.
The choice of the $CP$-eigenstates $|K_1\rangle$ , $|K_2\rangle$
is rather suited to experimental tests.
Indeed, since a two-pion state has the same $CP$ eigenvalue
of $|K_1\rangle$,
the probability $P_t(K_1)=\langle K_1|\rho(t)| K_1\rangle$ 
of having a kaon state
$K_1$ at time $t$ is directly related to the frequency 
of two-pion decays at time $t$.
Possible direct $CP$ violating effects, the only ones
allowing $K_2\to 2\pi$, can be safely neglected;
they are proportional to the phenomenological
parameter $\varepsilon'$, that is expected to be
very small, $|\varepsilon'|\leq 10^{-6}$.%
\footnote{The recent measures of ${\cal R}e({\varepsilon'/\varepsilon})$,
\cite{25},\cite{26} seem to confirm these
theoretical expectations.\cite{27}}

On the other hand, while the decay state
$\pi^0\pi^0\pi^0$ has $CP=-1$, the state 
$\pi^+\pi^-\pi^0$ may have $CP=\pm1$.
Thus, the probability 
$P_t(K_2)=\langle K_2|\rho(t)| K_2\rangle$ to have a kaon state 
$K_2$ at time
$t$ is not as conveniently  measured by counting the frequency of 
the three-pion decays.
To avoid the difficulty, the following strangeness eigenstates
can be used:
\begin{equation}
\label{24}
|K^0\rangle={|K_1\rangle+|K_2\rangle\over\sqrt{2}}\ ,\qquad 
|\overline{K^0}\rangle={|K_1\rangle-|K_2\rangle\over\sqrt{2}}\ .
\end{equation}
Then, the probabilities $P_t(K^0)=\langle K^0|\rho(t)|K^0\rangle$ and
$P_t(\overline{K^0})=\langle \overline{K^0}|\rho(t)|\overline{K^0}\rangle$,
that the kaon state at time $t$ be a $K^0$, 
respectively a $\overline{K^0}$, may be experimentally determined 
by counting the semileptonic decays $K_0\mapsto \pi^-\ell^+\nu$, 
respectively $\overline{K^0}\mapsto \pi^+\ell^-\overline{\nu}$, 
the exchanged decays being forbidden by the 
$\Delta S=\Delta Q$ rule. (In the Standard Model, this selection
rule is expected to be valid up to order $10^{-14}$.\cite{28})
Further, the probability $P_t(K_2)=\langle K_2|\rho(t)|K_2\rangle$
of having a kaon state $K_2$ at proper time $t$ can be expressed as
$P_t(K_2)=P_t(K^0)+P_t(\overline{K^0})-P_t(K_1)$, by writing
\begin{equation}
\label{25}
|K_2\rangle\langle K_2|=|K^0\rangle\langle K^0|+|\overline{K^0}\rangle
\langle\overline{K^0}|-|K_1\rangle\langle K_1|\ .
\end{equation}
Hence, $P_t(K_2)$ can be measured by counting the frequencies of semileptonic 
decays and of decays into two pions.

In order to measure the off-diagonal elements 
$\langle K_1|\rho|K_2\rangle$, $\langle K_2|\rho|K_1\rangle$, we use
the identity
\begin{equation}
\label{26}
|K^0\rangle\langle K^0|-|\overline{K^0}\rangle\langle \overline{K^0}|
=|K_1\rangle\langle K_2|+|K_2\rangle\langle K_1|\ ,
\end{equation}
and extract $|K_1\rangle\langle K_2|$ from it.
To do this, we need a third orthonormal basis of vectors 
whose projectors are measurable observables in actual experiments.
An interesting possibility is based on the phenomenon of kaon-regeneration
(see \cite{29} and references therein). 
The idea is to insert a slab of material across the neutral
kaons path; the interactions of the $K^0$, $\overline{K^0}$ mesons
with the nuclei of the material ``rotate'' in a known
way the initial kaon states entering the regenerator into new ones.
As initial states, consider the orthogonal vectors
\begin{equation}
\label{27}
|\widetilde{K}_S\rangle={|K_1\rangle-\eta^*|K_2\rangle\over\sqrt{1+|\eta|^2}}\ ,
\qquad
|\widetilde{K}_L\rangle={\eta|K_1\rangle+|K_2\rangle\over\sqrt{1+|\eta|^2}}\ ,
\end{equation}
where $\eta$ is a complex parameter which depends on the regenerating material.
By carefully choosing the material and the tickness of the slab,
one can tune the modulus and phase of $\eta$ in such a way
to completely suppress the $\widetilde K_L$ component and to
regenerate the $\widetilde K_S$ state into a $K_1$, just outside the
material slab. Thus, the probability 
$P_t(\widetilde{K}_S)=\langle \widetilde{K}_S|\rho(t)|\widetilde{K}_S\rangle$ 
that a kaon, impinging on a slab of regenerating material in a state $\rho(t)$ 
at time $t$, be a $\widetilde{K}_S$, can be measured by counting the decays 
into $2\pi$ just beyond the slab.
Now, the projector onto the state $|\widetilde{K}_S\rangle$ reads
\begin{eqnarray}
\nonumber
|\widetilde{K}_S\rangle\langle\widetilde{K}_S|&=&
{1\over 1+|\eta|^2}|K_1\rangle\langle K_1|+{|\eta|^2\over1
+|\eta|^2}|K_2\rangle\langle K_2|\\
\label{28}
&-&{\eta\over1+|\eta|^2}|K_1\rangle\langle K_2|-{\eta^*\over1+|\eta|^2}
|K_2\rangle\langle K_1|\ .
\end{eqnarray}
Then, from (\ref{24})--(\ref{26}) and (\ref{28}) it follows that
\begin{eqnarray}
\nonumber
|K_1\rangle\langle K_2|&=&f_1\, |K_1\rangle\langle K_1|\, +\, 
f_2\, |\widetilde{K}_S\rangle\langle \widetilde{K}_S|\\
\label{29}
&+&f_3\, |K^0\rangle\langle K^0|\, +\,
f_4\, |\overline{K^0}\rangle\langle \overline{K^0}|\ ,
\end{eqnarray}
where
\begin{eqnarray}
\label{30}
f_1&=&{1-|\eta|^2\over 2 i{\cal I}m(\eta)}\ ,\qquad  
f_2=-{1+|\eta|^2\over 2 i{\cal I}m(\eta)}\ ,\\
\label{31}
f_3&=&{|\eta|^2-\eta^*\over 2 i{\cal I}m(\eta)}\ ,\qquad
f_4={|\eta|^2+\eta^*\over 2 i{\cal I}m(\eta)}\ .
\end{eqnarray}
In this way, the determination of the off-diagonal elements of $\rho(t)$
amounts to counting the frequencies of decays into two pions with or without 
regeneration and the frequencies of semileptonic decays: 
\begin{eqnarray}
\nonumber
\langle K_1|\rho(t)|K_2\rangle&=&
f_1\, P_t(K_1)+f_2\, P_t(\widetilde{K}_S)\\
\label{32}
&+& f_3\, P_t(K^0)+f_4\, P_t(\overline{K^0})\ .
\end{eqnarray}

The application of these results to the case of correlated 
kaons is now straightforward. For sake of compactness, we
identify the various kaon states 
with the projections $Q_\mu$, $\mu=1,2,3,4$, where:
\begin{eqnarray}
\label{33}
&&Q_1=|K_1\rangle\langle K_1|\ ,\qquad\, 
Q_3=|K^0\rangle\langle K^0|\ ,\\
\label{34}
&&Q_2=|\widetilde{K}_S\rangle\langle\widetilde{K}_S|\ ,\qquad
Q_4=|\overline{K^0}\rangle\langle\overline{K^0}|\ .
\end{eqnarray}
As discussed, these operators can be measured by identifying
$2\pi$ final states, in absence and presence of a regenerator
($Q_1$ and $Q_2$), and semileptonic decays ($Q_3$ and $Q_4$);
the same holds for the projectors in (\ref{9}) and (\ref{10}),
since:
\begin{eqnarray}
\label{35}
P_1&=&|K_1\rangle\langle K_1|\equiv Q_1\ ,\\
\label{36}
P_2&=&|K_2\rangle\langle K_2|=Q_3 + Q_4 - Q_1\ ,\\
\label{37}
P_3&=&|K_1\rangle\langle K_2|\equiv P_4^\dagger=
\sum_{\mu=1}^4 f_\mu\, Q_\mu\ .
\end{eqnarray}
Further, we denote by $P_t(Q_\mu,Q_\nu)$ the 
probability that, at proper time $t$ after a $\phi$-decay, the two 
kaons be in the states identified by $Q_\mu$ and $Q_\nu$, respectively.
Then, the determination of the quantities 
${\cal P}_{ij}(t)$ reduces to 
measuring joint probabilities, {\it i.e.} to counting 
frequencies of events of certain specificied types.
Indeed, one explicitly finds:
\begin{eqnarray}
\label{38}
&&{\cal P}_{11}(t)=P_t(Q_1,Q_1)\ ,\\
\label{39}
&&{\cal P}_{12}(t)=P_t(Q_1,Q_3) + P_t(Q_1,Q_4) -
P_t(Q_1,Q_1)\ ,\\
\label{40}
&&{\cal P}_{13}(t)=\sum_{\mu=1}^4\, f_\mu\, P_t(Q_1,Q_\mu)\ ,\\
\nonumber
&&{\cal P}_{22}(t)=P_t(Q_1,Q_1)+P_t(Q_3,Q_3)+P_t(Q_4,Q_4)\\
\label{41}
&&\hskip 3cm +2\Big[P_t(Q_3,Q_4)-P_t(Q_1,Q_4)-P_t(Q_1,Q_4)\Big]\ ,\\
\label{42}
&&{\cal P}_{23}(t)=\sum_{\mu=1}^4\, f_\mu\, \Big[P_t(Q_3,Q_\mu)
+ P_t(Q_4,Q_\mu) - P_t(Q_1,Q_\mu)\Big]\ ,\\
\label{43}
&&{\cal P}_{33}(t)=\sum_{\mu=1}^4\sum_{\nu=1}^4\, 
f_\mu\,f_\nu\, P_t(Q_\mu,Q_\nu)\ .
\end{eqnarray}

As a result of the previous analysis, the inconsistencies of models without 
complete positivity, besides being theoretically unsustainable,
turn out to be experimentally exposable. 
Let $P_u$ and $P_v$ project onto the correlated states
\begin{eqnarray}
\label{44}
|u\rangle&=&{1\over\sqrt{2}}\Bigl(|K_1\rangle\otimes|K_1\rangle + 
|K_2\rangle\otimes|K_2\rangle\Bigr)\ ,\\
\label{45}
|v\rangle&=&{1\over\sqrt{2}}\Bigl(|K_1\rangle\otimes|K_2\rangle + 
|K_2\rangle\otimes|K_1\rangle\Bigr)\ .
\end{eqnarray}
The averages of these two positive observables with respect to the state
$\rho_A(t)$ read
\begin{eqnarray}
\nonumber
{\cal U}(t)&=&{\rm Tr}[\rho_A(t)\, P_u]\equiv\langle u|\rho_A(t)|u\rangle\\
\label{46}
&=&{1\over2}\bigl({\cal P}_{11}(t)
+{\cal P}_{22}(t)+{\cal P}_{33}(t)+{\cal P}_{44}(t)\bigr)\\
\nonumber
{\cal V}(t)&=&{\rm Tr}[\rho_A(t)\, P_v]\equiv\langle v|\rho_A(t)|v\rangle\\
\label{47}
&=&{1\over2}\bigl({\cal P}_{12}(t)
+{\cal P}_{21}(t)+{\cal P}_{34}(t)+{\cal P}_{43}(t)\bigr)\ ,
\end{eqnarray}
and, as explained before, can be directly obtained
by measuring joint probabilities in experiments at $\phi$-factories.
On the other hand, (\ref{14})--(\ref{20}) give, up to first order in the small
parameters,
\begin{eqnarray}
\nonumber
&&{\cal U}(t)=e^{-2\Gamma t}\Biggl[
{\gamma\over\Delta\Gamma}\sinh(t\Delta\Gamma)
+{b\over\Delta m}\Big(1-\cos(2 t\Delta m)\Big)\\
\label{48}
&&\hskip 7cm
+{a-\alpha\over2\Delta m}\sin(2 t\Delta m)\Biggr]\ ,\\
\label{49}
&&{\cal V}(t)=e^{-2\Gamma t}
\big(a+\alpha-\gamma\big)\, t\ .
\end{eqnarray}
Thus, ${\cal U}(0)={\cal V}(0)=0$, whereas
\begin{equation}
\label{50}
{{\rm d}{\cal U}(0)\over{\rm d}t}=a+\gamma-\alpha\ ,\qquad 
{{\rm d}{\cal V}(0)\over{\rm d}t}=a+\alpha-\gamma\ ,
\end{equation}
are both positive because of conditions (\ref{6}) and (\ref{7}).
More in general, the mean values (\ref{46}) and (\ref{47}) 
are surely positive, for the complete positivity of 
the single-kaon dynamical maps $\omega_t$ implies
$\rho(t)=\sum_j V_j(t)\, \rho\, V_j^\dagger(t)$,\cite{1}--\cite{3} 
where the $V_j(t)$ are $2\times 2$ matrices such that
$\sum_jV_j^\dagger(t)V_j(t)$ is a bounded 
$2\times 2$ matrix.%
\footnote{Notice that, in absence of the extra contribution
$L_D$ in (\ref{1}), the time evolution $\rho(t)$ is realized
with a single matrix $V$, {\it i.e.} $j=1$, and
$V_1(t)=e^{-i H_{\rm eff}\, t}$; in other words, in ordinary
quantum mechanics the condition of complete positivity is
trivially satisfied.}
Then, the complete evolution
$\rho_A\to\rho_A(t)=\sum_{i,j}[V_i(t)\otimes V_j(t)]\, \rho_A\,
[V_i^\dagger(t)\otimes V_j^\dagger(t)]$
will never develop negative eigenvalues.

On the other hand, if the single-kaon dynamical map $\omega_t$ 
is not completely positive, inconsistencies may emerge.
As an example, take the phenomenological 
models studied in \cite{21}, \cite{22},
where the non-standard parameters $a$, $b$, $c$ are set to zero and
$\alpha\neq\gamma$, $\alpha\gamma\geq\beta^2$. 
The corresponding dynamics is not completely
positive: the inequalities (\ref{5})--(\ref{8}) are in fact violated.
In this case, one still has ${\cal U}(0)={\cal V}(0)=0$, but 
\begin{equation}
\label{51}
{{\rm d}{\cal U}(0)\over{\rm d}t}=\gamma-\alpha=
-{{\rm d}{\cal V}(0)\over{\rm d}t}\ .
\end{equation}
Therefore, one of the mean values (\ref{46}), (\ref{47})
starts assuming negative values 
as soon as $t>0$.
The inconsistence is avoided only 
if $\alpha=\gamma$, which is a necessary condition for getting back
the property of complete positivity. We stress that
planned set-ups at $\phi$-factories can measure the two mean
values in (\ref{46}) and (\ref{47}) and therefore clarify
also from the experimental point of view the request of
complete positivity for the dynamics of neutral kaons.

\end{document}